\documentclass[11pt, preprint]{aastex}
\usepackage{geometry}                
\usepackage{graphicx}
\usepackage{amssymb}
\usepackage{natbib}
\usepackage{amsmath}
\usepackage{epstopdf}
\newcommand{\bx}{{\bf x}}

\newcommand{\bnabla}{{\boldsymbol \nabla}}
\newcommand{\bxi}{{\boldsymbol \xi}}
\newcommand{\bu}{{\boldsymbol v}}

\begin{document}
\title{Full Waveform Inversion for Time-Distance Helioseismology}
\author{Shravan M. Hanasoge\altaffilmark{1,2,3} \& Jeroen Tromp\altaffilmark{2,4}}
\altaffiltext{1}{Department of Astronomy and Astrophysics, Tata Institute of Fundamental Research, Mumbai 400005, India}
\altaffiltext{2}{Department of Geosciences, Princeton University, Princeton, NJ 08544, USA}\email{hanasoge@tifr.res.in}
\altaffiltext{3}{Max-Planck-Institut f\"{u}r Sonnensystemforschung, 37191 Katlenburg-Lindau, Germany}
\altaffiltext{4}{Program for Applied and Computational Mathematics, Princeton University, Princeton, NJ 08544, USA}

\begin{abstract}
Inferring interior properties of the Sun from photospheric measurements of the seismic wavefield constitutes the helioseismic inverse problem.
Deviations in seismic measurements (such as wave travel times) from their fiducial values estimated
for a given model of the solar interior imply that the model is inaccurate.
Contemporary inversions in local helioseismology assume that properties of the solar interior are linearly related
to measured travel-time deviations. It is widely known, however, that this assumption is invalid for sunspots and
active regions, and likely for supergranular flows as well.

Here, we introduce nonlinear optimization, executed iteratively, as a means of inverting for the sub-surface structure of large-amplitude
perturbations. Defining the penalty functional as the $L_2$ norm of wave travel-time deviations, we compute the
the total misfit gradient of this functional with respect to the relevant model parameters 
at each iteration around the corresponding model. The model is successively improved using either steepest descent, conjugate gradient, or
quasi-Newton limited-memory BFGS. Performing nonlinear iterations requires privileging pixels (such as those in the near-field of the scatterer), a practice
not compliant with the standard assumption of translational invariance. Measurements for these inversions,
although similar in principle to those used in time-distance helioseismology, require some retooling.
For the sake of simplicity in illustrating the method, we consider a 2-D inverse problem with only a sound-speed perturbation.
\end{abstract}

\section{Introduction}
Imaging the non-axisymmetric interior structure and dynamics of the Sun requires interpreting measurements of the photospheric
seismic wavefield \citep[see reviews by, e.g.,][]{gizon05, gizon2010}. There exist a number of techniques to process observations of the seismic wavefield; 
in this article we focus on time-distance helioseismology \citep{duvall}, in which travel times of waves are the primary measurements. 

Full waveform inversion is a label for set of techniques widely used in terrestrial and exploration seismology to infer the structure of the highly
heterogeneous Earth. ``Full waveform" refers to the use of the entire seismic measurement (which in
the case of helioseismology is the cross correlation) in the inversion. A waveform can broken up into frequency bands,
and every part of the waveform can be characterized by parameters such as phase and amplitude. The full-waveform approach
involves assimilating all of these measurements into the inversion in the to maximally leverage seismic data. A number of inversion
methods already adopt aspects of this approach \citep[e.g.,][]{svanda11, jason12, dombroski13}, strictly assuming however that
seismic measurements depend linearly on interior properties.
{In the present formulation, we compare waveforms
solely in the sense of travel times. Further, because we only consider sound-speed perturbations here,
the primary impact on waveforms is to shift their phases and to a lesser degree, amplitude.
In principle, we may also include amplitudes, instantaneous phase, or even raw
waveform differences \citep[e.g.,][]{dahlen02,bozdag11,rickers2013}. }


The basic goal  
in seismology is to relate properties of the interior to wavefield measurements at the bounding surface.
The first step involves defining a misfit or cost functional that comprises some measure of the difference between measurement and prediction.
An example of a misfit function ($\chi$) in the case of time-distance helioseismology is the $L_2$ norm of the difference between measurement ($\tau^{\rm o}$) and prediction
($\tau$) at some set of locations $i$ \citep{hanasoge11}
\begin{equation}
\chi = \frac{1}{2}\sum_i (\tau_i -\tau^{\rm o}_i)^2. \label{misf}
\end{equation}
{A more general formulation to include a noise-covariance matrix in the definition of the misfit is discussed by \citet{hanasoge11}. Here, we study
a simpler problem where the data are known exactly.}
The next step is to determine how to change the model so that the predicted travel times $\tau_i$ are closer to the measurements $\tau^{\rm o}$ in the 
sense of norm~(\ref{misf}). 
This is a high-dimensional inverse problem, since we seek to alter various 
properties such as flows, sound speed 
and density of the 3-D interior, thereby introducing a large number of parameters, in order to appropriately alter the travel times measured at the bounding surface of the Sun. 

The misfit function~(\ref{misf}) depends on the model, i.e., $\chi = \chi({\bf m})$, where ${\bf m} = {\bf m}(\bx)$ is the model of the Sun and $\bx$ is the spatial coordinate.
To vary the misfit, we consider the Taylor expansion of equation~(\ref{misf}) around model ${\bf m}$,
\begin{equation}
\delta\chi = \sum_i (\tau_i -\tau^{\rm o}_i)\,\frac{\partial\tau_i}{\partial{\bf m}}\,\delta{\bf m},\label{taylor}
\end{equation}
and it is seen that to reduce the misfit, i.e., to induce $\delta\chi < 0$, we first need access to the gradient of the misfit function ${\partial\tau_i}/{\partial{\bf m}}$.
Gradient-based optimization methods
are designed to address this question, specifically to minimize penalty~(\ref{misf}), an inherently non-linear function of the 3-D model parameters.
The gradient of misfit~(\ref{misf}) with respect to model parameters is the so-called `sensitivity kernel', alternately known as the Fr\'{e}chet derivative,
\begin{equation}
\frac{\partial\tau_i}{\partial{\bf m}} = {\bf K}(\bx,\bx_i; {\bf m}),\label{kern.exp}
\end{equation}
where ${\bf K}$ is the sensitivity of travel time $\tau_i$ to changes in the model ${\bf m} = {\bf m}(\bx)$, and is therefore a function of the model and space. Equation~(\ref{kern.exp}) along with~(\ref{taylor}) gives us a prescription to compute a model that minimizes the misfit for the quiet Sun,
\begin{equation}
\delta\chi = \int_{\odot} d\bx\,K_c\, \delta \ln c + K_\rho\, \delta \ln\rho + {\bf K}_{\bu}\cdot \delta\bu,\label{delchi}
\end{equation}
where $c$ is sound speed, $\rho$ is density and $\bu$ are flows, $K_c, K_\rho$, and ${\bf K}_\bu$ are kernels for sound speed, density and flows respectively \citep{hanasoge11, hanasoge12_mag}. We use log quantities
for variations in $c$ and $\rho$ since they are positive definite. 

This article aims to introduce the basic concepts of this inverse methodology and is not exhaustive in its scope. We therefore limit
ourselves to the study of a sound-speed inversion, described thus
\begin{equation}
\delta\chi = \int_{\odot} d\bx\,K_c\, \delta \ln c.\label{delc}
\end{equation}
To compute the misfit gradient $K_c$, we apply the adjoint method described by \citet{hanasoge11},
used to simultaneously construct kernels $K_c, K_\rho,$ and ${\bf K}_\bu$ \citep[also see e.g.,][]{tarantola84, tromp05}. However, we only retain $K_c$ for this problem.

Seismic inversions are matrix-inverse problems of the form
\begin{equation}
A\,\delta{\bf m} = \{\delta\tau\},\label{matinv}
\end{equation}
where $A = A({\bf m})$ is a fat matrix of dimension $N\times M$, and where the $M$ unknown model parameters are substantially larger 
than the $N$ measurements, $\delta{\bf m}$ is the model update vector, of size $M\times1$ and $\{\delta\tau\}$ is an $N\times1$ vector 
composed of the travel times. The matrix $A$ comprises the sensitivity of the travel time to model parameters, i.e., it is composed of
sensitivity kernels. At present, inverse problems in local helioseismology focus on constructing sensitivity kernels using only 1-D vertical stratification,
{leading to lateral (horizontal) translation invariance}. 
Although likely erroneous {for certain problems}, this approach is generally invoked
regardless because a viable methodology to fully account for the three-dimensionality and non-linearity of the inverse problem has only
recently been introduced \citep{hanasoge11}. 
{Inverse approaches that rely on translation invariance possess the additional feature that the computational cost
scales very weakly with the number of measurement points, unlike in the adjoint method. On the other hand, it is possible to mitigate
the computational cost of adjoint-method based approaches by choosing a set of observation points such that coverage and resolution
are maximized.}

Matrix $A$ can be very big (with $10^{12}$ elements or more), and will possess a high condition number, and therefore inverting it is not an option. Consequently,
we use an iterative procedure to arrive at some appropriate inverse of $A$ and therefore, $\delta{\bf m}$. To perform iterations, a local linear approximation is invoked,
much as in the style of the Taylor expansion in equation~(\ref{taylor}), and methods such as steepest descent, conjugate gradient
or the quasi-Newton limited-memory BFGS are applied.

The adjoint method, a means of obtaining gradients of the misfit function $\chi$, is well studied in the regime of relatively strong heterogeneities,
as demonstrated by the successful application to terrestrial seismic inversions of, e.g., the Southern-California crust \citep{CarlTape09}, European
structure \citep{zhu_attenuation} and Australia \citep{fichtner09}.
This technique is applied to constrained optimization problems in which we seek to minimize the misfit with the constraint that the wavefield satisfy the
partial differential equation that governs wave propagation in the Sun.
We define the helioseismic operator,
\begin{equation}
\rho\partial^2_t\bxi = \bnabla(\rho c^2\bnabla\cdot\bxi + \rho g\xi_z) +  {\bf g}\,\bnabla\cdot(\rho\bxi) + {\bf S},\label{goveq}
\end{equation}
where density is denoted by $\rho = \rho(\bx)$, sound speed by $c = c(\bx)$, gravity by ${\bf g} = - g(z)\, {\hat{\bf z}}$, the vector acoustic wave displacement by $\bxi = \bxi(\bx,t)$, whose vertical component is $\xi_z$, the source by ${\bf S} = {\bf S}(\bx,t)$ and time by $t$. The covariant
spatial derivative is denoted by $\bnabla$ and the partial derivative with respect to time is $\partial_t$. 
The adjoint method relies on making predictions and using the difference with observations to drive changes in the solar model. Thus,
we require a technique to solve equation~(\ref{goveq}). The pseudo-spectral solver {\rm{SPARC}} developed by \citet{dealias, Hanasoge_couvidat_2008},
fulfills the purpose of solving equation~(\ref{goveq}) in Cartesian geometry.  
Lateral (horizontal) derivatives are computed using Fourier transforms and the radial (vertical) derivative using a sixth-order accurate compact-finite-difference scheme \citep{lele92}.
Time stepping is achieved through the repeated application of an optimized second-order five-stage Runge-Kutta technique \citep{hu}. We line the side and vertical
boundaries with perfectly matched layers \citep{hanasoge_2010} that effect high fidelity wave absorption.

The adjoint method consists of computing {\it forward} and {\it adjoint} wavefields.
The {\it forward} calculation is a predictor step, making a prediction on the
photospheric cross correlation (or some other measurement) along with the attendant 3-D seismic wavefield in the interior.
This calculation captures the connection between the interior sensitivity of the wavefield and the surface seismic signature.
The {\it adjoint} calculation consists of performing a 3-D wavefield simulation driven by the 
difference between prediction and observation, as measured by equation~(\ref{misf}). Roughly speaking, this captures the connection
between the interior and the measurement misfit as recorded at the surface. 
Finally, the time-domain convolution of forward and adjoint wavefields gives the total misfit gradient, i.e., 
all the desired sensitivity kernels (Eq.~[\ref{delchi}]). 
Because this formulation of the adjoint method is numerical, forward and adjoint simulations may be carried out for arbitrary backgrounds.
Further, with a few calculations, all relevant kernels may be simultaneously obtained. The analysis, kernel expressions and algorithm are outlined in sections 4, 5 and 6 
respectively of \citet{hanasoge11}. 
Finally, we note that the extension to a variety of other measurements such as resonant frequencies closely follows the analysis in section 4 of \citet{hanasoge11}, with
the relevant measurement framed in a manner so as to connect it to Green's functions of the medium.

Waves in the Sun are excited in a thin near-surface radial envelope \citep[e.g.,][]{stein00} but uniformly in the lateral (horizontal) direction. Thus
the helioseismic wavefield is excited by distributed sources, which, together with the stochastic nature of the excitation, makes the calculation
of sensitivity kernels complicated \citep[][]{hanasoge11}. 
This is because the wavefield measured at a given point consists of contributions from a wide range of sources and the cross-correlation of
the wavefield measured at a point pair thus averages these contributions.
However, in the case where the distribution of sources is uniform, the cross-correlation can be shown to be closely related to Green's function of
the medium \citep[e.g.,][]{snieder04}. This correspondence allows for treating the second-order cross-correlation measured between a point pair as arising from
a deterministic, single source-receiver configuration, greatly reducing the complexity of the problem (the point pair map on to the source and receiver). 
While it may appear that the solar wavefield is an ideal fit for this correspondence (owing to the lateral uniformity of sources), the damping
mechanism and the line-of-sight nature of observations diminish the accuracy of the relationship \citep[e.g.,][]{gizon2010}.
However, it still serves as a very useful first
approximation to study the simplified deterministic source-receiver problem since it
allows for the appreciation and development of inverse methodology prior to comprehensive modeling.
Kernels in this limit treat each branch of the cross correlation measured between a pair of points as the wave
displacement due to a deterministic single source. 

\section{The inversion}
The road to obtaining consistent inversions is long, requiring a number of important steps to be implemented. Here we discuss
practical issues and the choices we have made. We do not start from a vacuum, and indeed, there exists significant
geophysical seismic literature on these topics, and the choices from these articles guide our thinking. However, the
helioseismic inverse problem possesses its own idiosyncrasies and to optimize our methodology, an exhaustive survey of these choices will be necessary.
This is especially the case when including more parameters such as flows and magnetic fields. 

\subsection{True and starting models}
The goal is to invert for the true anomaly in sound speed shown in Figure~\ref{true}. 
Also shown in Figure~\ref{true} is the starting model, which is a solely vertically stratified, convectively stabilized 
form of model S \citep{jcd, hanasoge_thesis, Hanasoge_couvidat_2008}. Sound-speed perturbations shown in Figure 
are measured as deviations from this `quiet Sun' stratification, i.e., $[c(x,z) - c_q(z)]/c_q(z)$, where $c_q$ is the nominal
sound-speed in the quiet Sun and $c(x,z)$ is the sound speed of the current model. 
To accelerate convergence, we may also constrain the surface layers in the starting model to be identical to those of
the true model, the argument being that the surface layers of the true model would be `observable' (which we do in Section~\ref{const.surf}). 
For now, we choose the starting model, $c(x,z) = c_q(z)$. In the subsequent discussion and in various Figures and attendant 
captions, we will make use of the following definition
\begin{equation}
\delta\ln c = \ln{\frac{c(x,z)}{c_q(z)}}.\label{deltc}
\end{equation}

\subsection{Master and slave pixels}
\begin{figure}[!ht]
\begin{centering}
\epsscale{1}
\plotone{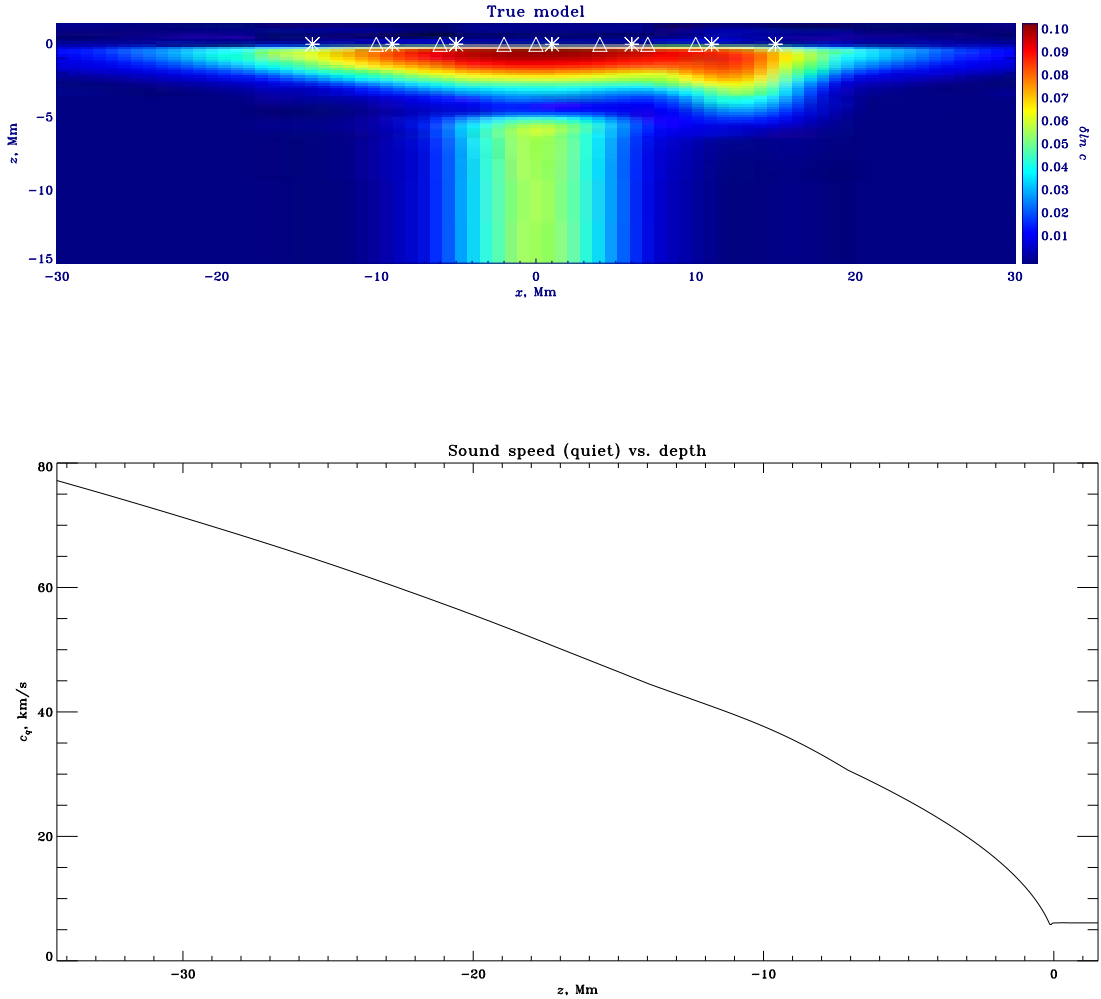}
\caption{True model (upper panel), where $\delta\ln c$ is defined in equation~(\ref{deltc}), and the quiet-Sun sound-speed, $c_q(z)$ in the lower panel.
The triangles denote the first set of master pixels (sources) and the asterisks the second set. The master pixels are switched at iteration 7, to 
introduce new seismic information.
Because wave excitation occurs in the very near-surface layers of the Sun ($z = -50$ km), we fix
the location in depth but are free to vary the horizontal location.
\label{true}}
\end{centering}
\end{figure}

Recalling the discussion on source-receiver pairs in the preceding section, we term
sources as master pixels and receivers as slaves.
\citet{tromp10} and \citet{hanasoge11} showed that the cost of inversion scales with the number of master pixels and hence
the nomenclature. Thus having selected points at which to place sources (master pixels),
we may increase the number of receivers (slaves) arbitrarily without accruing additional computational cost. Choices for
master pixels are therefore crucial since we would like to maximize seismic information. There are likely more formal
and rigorous ways to make this choice but in the effort here, we have discovered through the process of trial and error that
placing master pixels in the near field of the perturbation leads to faster convergence. We thus choose 7 master pixels
placed at points along the sound-speed perturbation as shown in Figure~\ref{true}. In order to introduce more seismic information,
we perform a few iterations for a given set of master pixels and replace these by another set. In the inversion presented here,
the master pixels change from the originally chosen set (indicated by triangles in Figure~\ref{true}) to another set of 7 pixels 
at iteration 7, indicated by asterisks. The new set of pixels is more sparsely distributed and is spread out over a larger
horizontal distance, to improve the imaging aperture. We do not introduce further changes
to the set of masters because seismic information is concentrated in the vicinity of the perturbation, which we
explore thoroughly with the overall set of pixels. Slave pixels may also be changed from iteration
to iteration, but here, we have maintained the same set of receivers throughout the inversion.

\subsection{Measurements}
\begin{figure}[!ht]
\begin{centering}
\epsscale{1}
\plotone{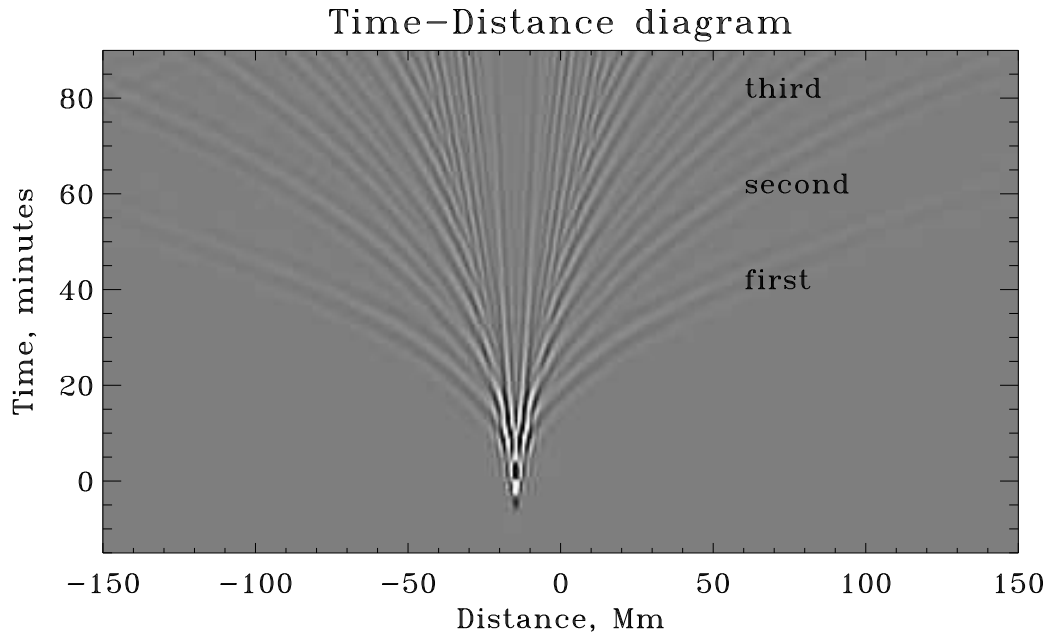}
\caption{Time-distance diagram. The master pixel (source) in this case is placed at $x = -15$ Mm. Travel time shifts measured at 
slave pixels (receivers) for a given bounce (first, second, or third) are used in the inversion.
In order to distinguish between the various arrivals, we select receivers that are at a minimum distance of 15 Mm away from the source 
for the first and second bounces and 30 Mm for the third bounce. 
\label{td}}
\end{centering}
\end{figure}

We measure wave travel times between point pairs. Using the definition of the linear travel time as
set out by \citet{gizon02}, we formulate the adjoint method for this measurement \citep{hanasoge11}.
In practice, the relative travel time between two waveforms is measured by actually cross correlating them and extracting the time lag
associated with the peak correlation coefficient.  For instance, if waves appear at point B at a positive
time lag in relation to point A, then point B acts as the receiver (slave) to source A (master). In Figure~\ref{td}, we show
the time-distance diagram for a source at $x=-15$ Mm.
 We measure travel times for $p$ modes over a range of point-pair distances for the first, second
and third bounces over specified frequency bands. 

\subsection{Adjoint source}
\begin{figure}[!ht]
\begin{centering}
\epsscale{1}
\plotone{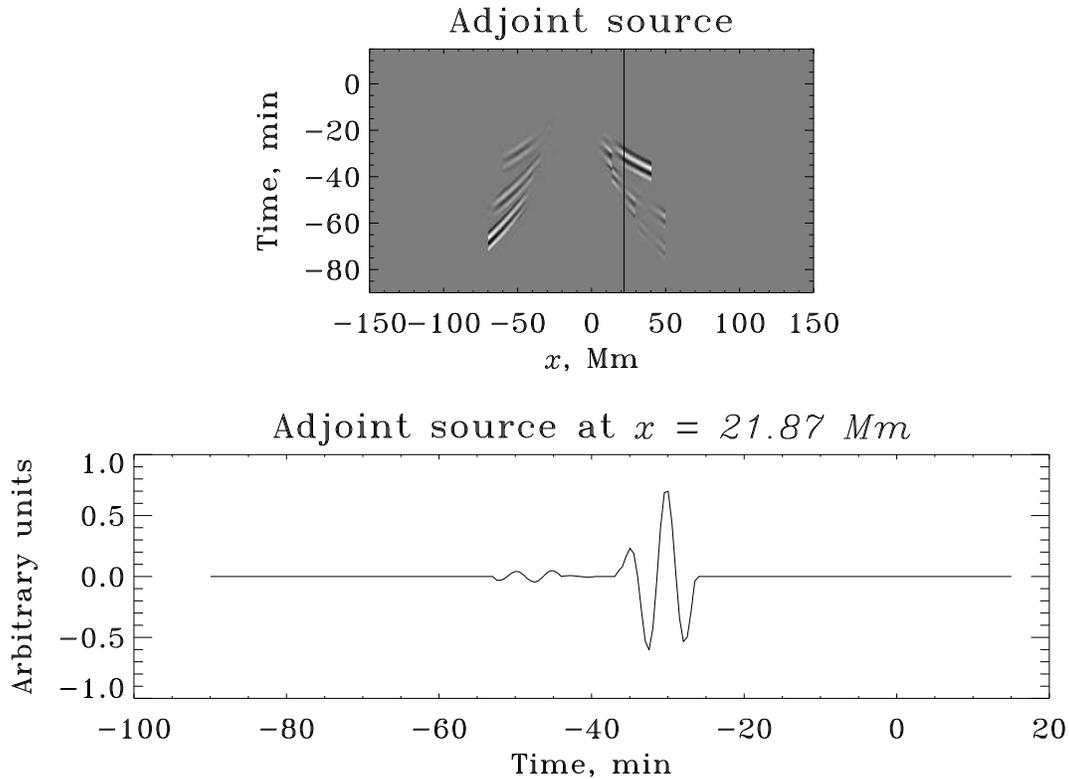}
\caption{Adjoint sources at receivers (upper panel) corresponding to the master pixel shown in Figure~\ref{td}. Each adjoint source is the time-reversed temporal derivative
of the waveform measured at that receiver, multiplied by the cross-correlation travel time shift. The adjoint source at a specific $x$ location is shown in the lower
panel. The waveform multiplied by the travel-time shift is the largest for the first bounce, which, owing to time reversal, appears at a later time in the adjoint source.
The adjoint source suggests that the most significant travel-time deviations are recorded by the first bounce, thereby playing a prominent role when constructing the gradient.
\label{adjointsource}}
\end{centering}
\end{figure}

For a given source point, we measure travel times at receivers located farther than 15 Mm from it. This minimum separation allows
for the distinction between the various bounces of $p$ modes. At distances shorter than 15 Mm, it is no longer possible to clearly
interpret the measurement. We only simulate for 1.5 hrs of solar time, which places a restriction on a maximum source-receiver distance possible for each
bounce. In the adjoint calculation, the wave equation is forced with {\it adjoint sources} placed at all the receiver locations where measurements are made.
The adjoint source at any given measurement point consists of the travel-time shift multiplied by the time reverse of the temporal derivative of the measured
 waveform from the forward calculation. In Figure~\ref{adjointsource}, the full adjoint source is shown in the upper panel and a cut at a fixed spatial location 
 is shown in the bottom.

\subsection{Discrete adjoint method}
In the formulation adopted here, the adjoint method is treated in a continuous sense \citep{hanasoge11}, and expressions for kernels that are computed
by convolving the forward and adjoint wavefields are derived for continuous space. However, numerical simulations are performed on discrete
grids, and indeed, errors are generated when the continuous adjoint formulation is discretized. The gradient thus obtained is not
as accurate as when the problem is posed consistently in the discrete sense. This slows down convergence and is a well noted issue in
these seismic inverse problems \citep[for airfoil design, see e.g., ][]{giles.adjoint.00}. Nevertheless, because convergence is observed and because there is no easy or obvious route to 
a discrete adjoint formulation, we proceed with the (inaccurate) continuous analog.
 
 \subsection{Preconditioning and Smoothing}
 While adjoint methods may not explicitly state the role of regularization, it does
 make its way into the heart of the problem. At every iteration, the total misfit gradient, summed over all master pixels, contains
 non-smooth variations co-spatial with source locations, which may slow convergence. 
 To mitigate this problem, spatial smoothing must be applied to the gradient. 

 The rate of convergence can be improved by `preconditioning' the gradient, which in practice involves multiplying the gradient by a suitable function termed the preconditioner,
 i.e., the gradient is preconditioned first and spatially smoothed next.
 The sensitivity of the convergence rate to different types of preconditioners was studied by \citet{tromp13},
 who found that the optimal preconditioner for the problem they were studying was 
 a convolution of the time derivatives of the forward and adjoint wave fields (see their Eqs.~[108] and [109]).
 However, we found that preconditioning \citep[based on the methods of][]{tromp13} and smoothing led to slower convergence rate
 in comparison to just smoothing. The design and application of preconditioners to helioseismology is deferred to the future and 
 we restrict ourselves only to smoothing the gradient here. 
Note that explicit regularization terms (user prescribed) may indeed be included in the original statement of the problem, since the adjoint method is
designed to address constrained-optimization problems.

\begin{figure}[!ht]
\begin{centering}
\epsscale{1}
\plotone{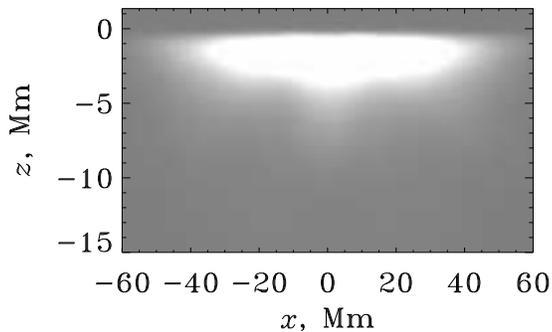}
\caption{The raw sound-speed gradient, shown in the upper panel has sharp variations due to numerical issues related to the spatially localized forward source.
The smoothed kernel is shown in the lower panel, where a 3-point Gaussian filter was applied to accomplish smoothing. 
The update is then computed through $c_{02} = c_{01}(1 + \varepsilon\bar{K_{c_{01}}})$, where the overbar indicates smoothing, $c_{02}$ is the sound-speed model for the second iteration and $\varepsilon$ is a small constant.
\label{smoothing}}
\end{centering}
\end{figure}

 \subsection{Model updates}
Given the gradient, the model can be updated using a variety of methods. The first iteration relies on steepest descent, in which
the update is tangent to the gradient direction. At higher iterations, we may choose between conjugate gradient and L-BFGS to
create updates. Conjugate gradient requires the previous and current gradients to form the update where L-BFGS can be designed
to use the full history of gradients and models to create an update. Although not shown here, from preliminary testing we find that
L-BFGS and conjugate gradient converge at roughly the same rate. More careful testing may reveal the parameter regimes where
one method is faster than the other.

\begin{figure}[!ht]
\begin{centering}
\epsscale{0.5}
\plotone{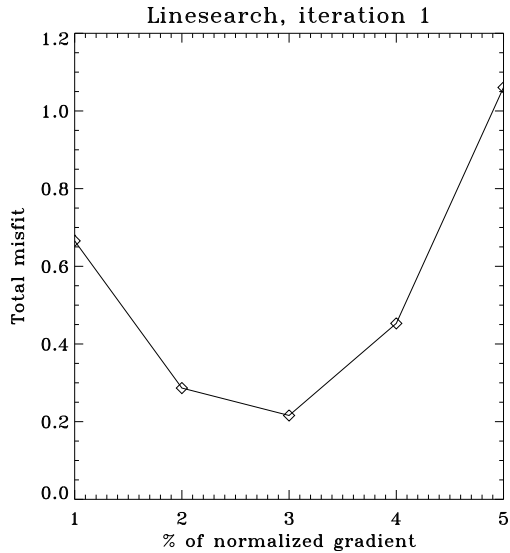}
\caption{Line search at each iteration to determine $\varepsilon$ for the update $c_{i+1} = c_{i}(1 + \varepsilon\,\bar{K_{c_i}})$. The $x$ axis
shows different values of $\varepsilon$ and the $y$ axis the misfit associated which the corresponding model. In this case, we choose the model
for which the misfit reaches a minimum, i.e., for $\varepsilon = 0.03$. 
\label{linesearch}}
\end{centering}
\end{figure}

Since we only consider sound-speed perturbations, the smoothed sound speed sensitivity kernel is first 
normalized by its largest absolute value so that it ($\bar{K_{c_i}}$) spans the range $[-1, 1]$. We then perform a line search, using 5 different models,
$c_{i+1} = c_{i}(1 + \varepsilon\,\bar{K_{c_i}})$, where $c_{i}$ is the model at the $i$th iteration, $\varepsilon$ is a small constant that takes on
values $[0.01, 0.02, 0.03, 0.04, 0.05]$. Every value of $\varepsilon$ leads to a model $c_{i+1}$, and we estimate the misfit for each. 
At every iteration, we test for local convexity by performing a line search. Typically an elegant $L$-curve is observed, as in Figure~\ref{linesearch}. 
We choose the model corresponding to the minimum point of this curve as the model for the next iteration, i.e. the update corresponds
to the valley of the line search curve.
The update parameter $\varepsilon$ generally decreases with iteration, and $\varepsilon$ for updates to successive models is smaller in magnitude.
Typically, $\varepsilon\sim 0.06$ for the very first iteration and then drops to about $\varepsilon \sim 0.004$ at the eleventh iteration.

\begin{figure}[!ht]
\begin{centering}
\epsscale{1}
\plotone{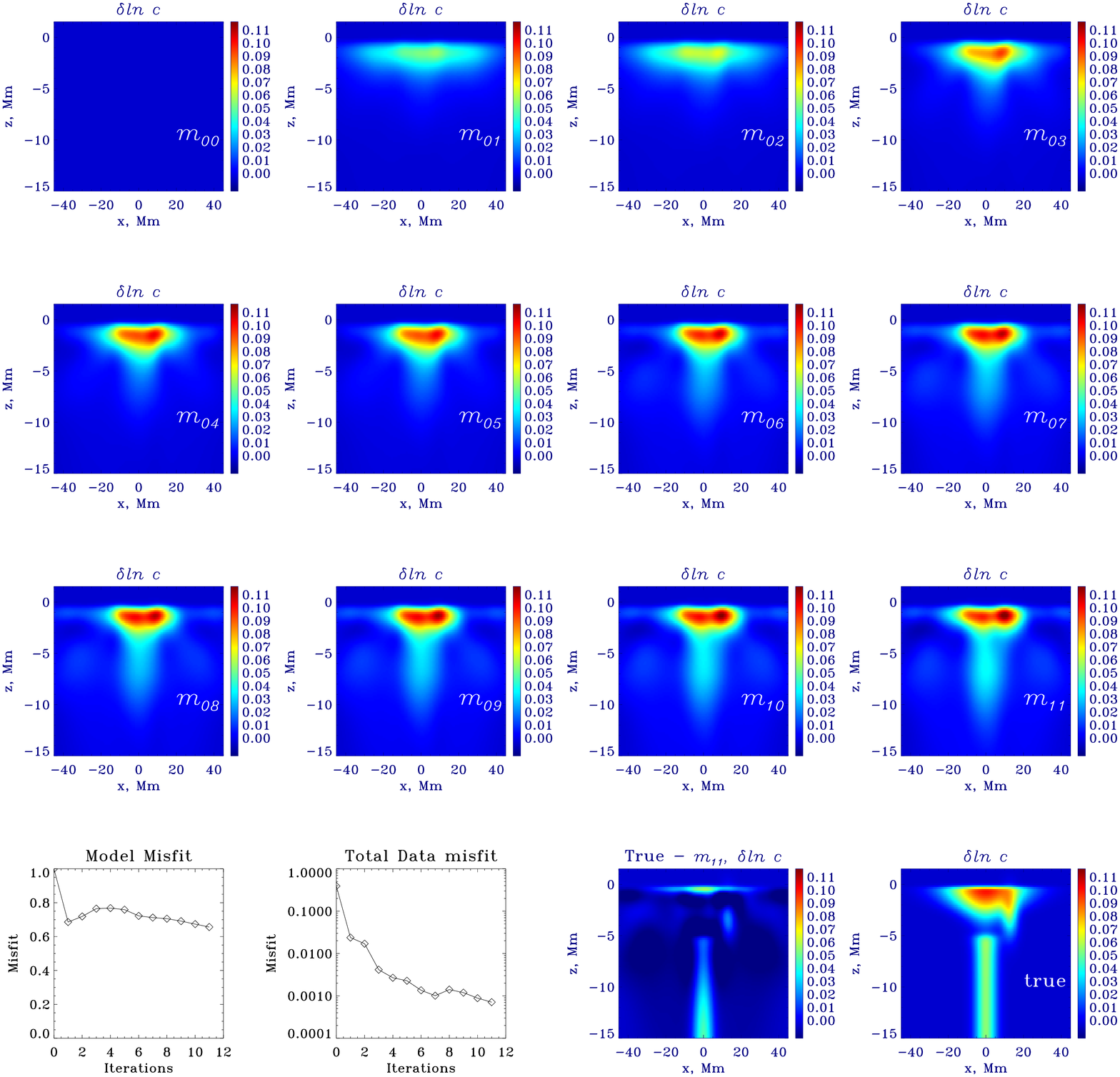}
\caption{Iterations in a conjugate-gradient based inversion. The first iteration is performed using steepest descent and a combination
of conjugate gradient and steepest descent are used to compute subsequent models. At iteration 7, we change the set of master pixels and
this creates a local jump in the data misfit because more information has been introduced. It is seen that models approach the
true anomaly gradually but the reduction in both data and model misfits slows down with iteration. 
The model misfit is the normalized $L_2$ norm difference between 
the true and current model whereas the total data misfit is the same as equation~(\ref{misf}). In the first few iterations, the model misfit
increases because surface layers contain significant errors and $p$ modes possess limited sensitivity to these layers. As the model evolves
it overcomes this local hill, appearing to `fix' the surface layers, and a steady decline is seen in the last few iterations.
\label{modelsconj}}
\end{centering}
\end{figure}

Every few iterations, the $L$-curve for a non-steepest-descent method is not easily produced. In such scenarios, we revert to
steepest descent as a means of `resetting' the inversion. For instance, we might have the following configuration of updates - 1 - steepest, 2, 3, 4 - conj. grad., 5 - steepest, 6, 7 - conj. grad,
where the numbers indicate the iteration index. 
We show 12 iterations of an inversion for the setup discussed in Figure~\ref{true} using a combination the conjugate gradient
and steepest descent methods in Figure~\ref{modelsconj}. We also applied the L-BFGS algorithm after 4 iterations of steepest descent
but found the rate of convergence to be generally unchanged. The performance of the method appears to be less sensitive to these
choices and much more to the introduction of external information (such as surface constraints, new pixels etc.).
  
\subsection{Uniqueness}
In high-dimensional inverse problems, the choice of the starting model and type of measurements introduced to update
the model may be critical to avoiding being trapped in a local minimum. 
A standard strategy applied to mitigate the chances of encountering this undesirable outcome is to first use measurements taken
from low frequency modes and gradually introduce higher frequencies as the model iteratively accrues features. This particular
issue can be very serious when attempting to image reflectors in the interior, as in exploration geophysics, but it is unlikely to be critical
for helioseismology. Because the frequency range of trapped modes in the Sun is so narrow (2.5 - 5.5 mHz), we choose here
to utilize the entire passband. Indeed, we are aware that this strategy may not be optimum for all applications but we find it to
be successful in the case of sound-speed perturbations studied here. 

\subsection{Testing convergence}
\begin{figure}[!ht]
\begin{centering}
\epsscale{1}
\plotone{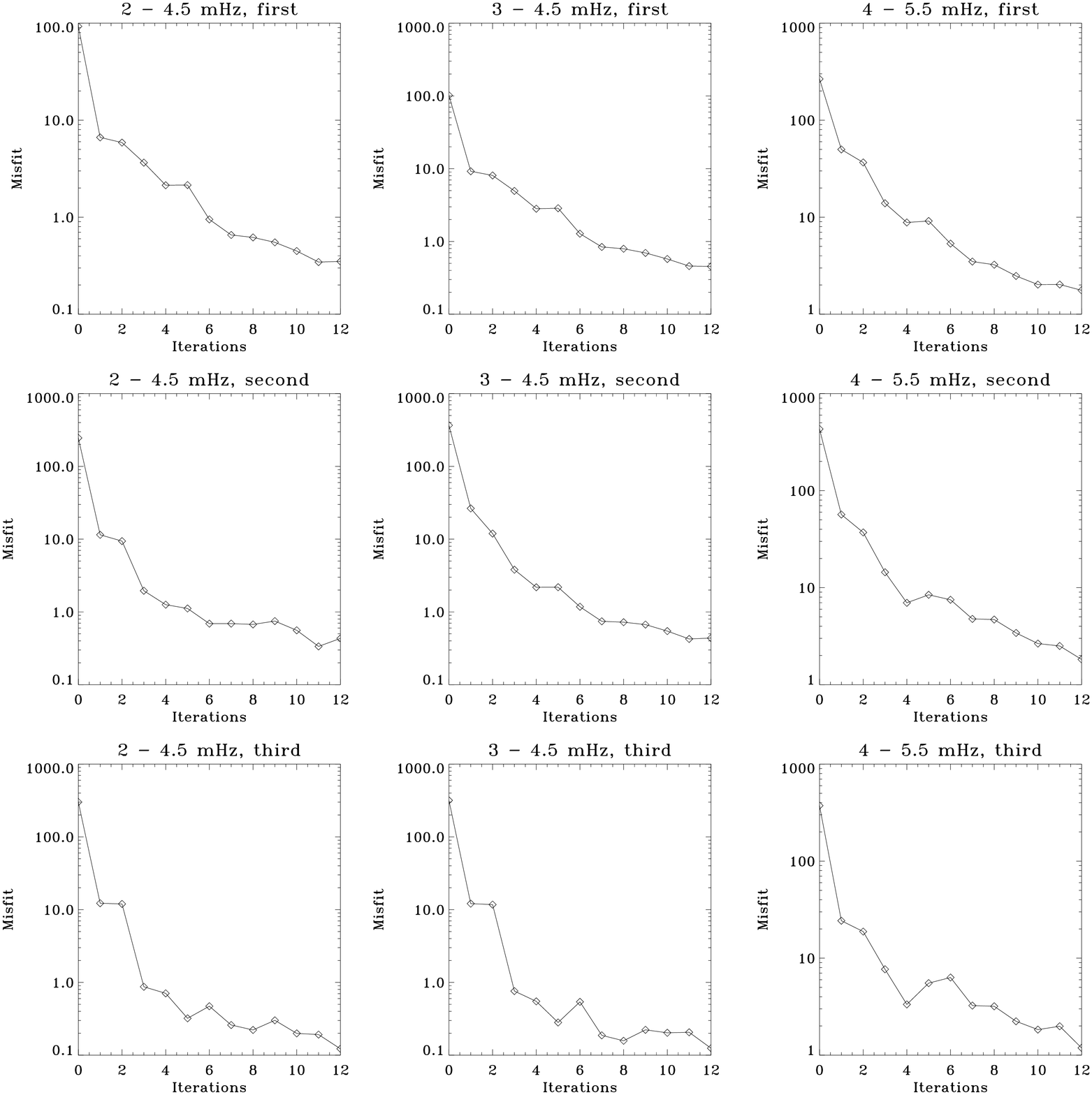}
\caption{Misfit reduction with iteration, broken up based on the frequency bands and bounces. It is seen that regardless of
the band, the misfit decreases uniformly (straying from monotonic reduction along the way on a few occasion). 
Note that we do not apply a frequency filter in our travel-time measurements, so we are not explicitly attempting to minimize
these separate bands. This trend occurs organically, suggesting that the eventual result will be consistent with the governing wave equation
and the measurement technique. It also adds support
to the notion that the adjoint method in conjunction with linear algebraic inverse methods can be very successful. Note that we could
also have used ridge- and phase-speed filtering to further test for a decreasing misfit with iteration.
\label{misfitband}}
\end{centering}
\end{figure}

To verify that misfit is being minimized
for all the measurements, we measure the misfit associated with each model for travel times binned into categories by their bounce 
number (first, second or third) and frequency band (2.5 -- 4, 2.5 -- 5, 2.5 -- 5.5). Note that we could also have measured the misfit using ridge- and phase-filtering
to isolate modes in various parts of the power spectrum but our categories are simpler in this case. 
Thus we confirm that the misfit is uniformly reduced
in these 9 categories. A similar strategy has been used successfully in terrestrial applications, e.g., \citet{zhu_attenuation} although because terrestrial
seismic waves exhibit a larger temporal frequency range, they apply frequency filters to their data. Fixing the lower frequency cutoff,
\citet{zhu_attenuation} increase the upper corner of the bandpass with iteration, gradually allowing in more information as the model
grows in complexity. We also calculate the model misfit by computing the $L_2$ norm of the difference between the true and 
inverted models as a function of iteration. Both data and model misfit are seen to decrease with iteration in Figure~\ref{misfitband}.

\subsection{Including ``surface" constraints}\label{const.surf}
The sound-speed anomaly studied here has a `surface' signature and we can include this as a constraint on the model.
It is of relevance because in reality, perturbations such as supergranules, meridional circulation, sunspots and active regions are
optically observed at the photosphere and these observations can be used to accelerate convergence. 
For the inverse problem at hand, $p$ modes are used to image the sound-speed perturbation.
Surface-gravity $f$ modes, which are very sensitive to the surface, do not register sound-speed perturbations since
the restoring force for these waves is gravity and not pressure. Consequently, adding a surface constraint to the inversion is likely to accelerate
convergence for this inverse problem.

\begin{figure}[!ht]
\begin{centering}
\epsscale{1}
\plotone{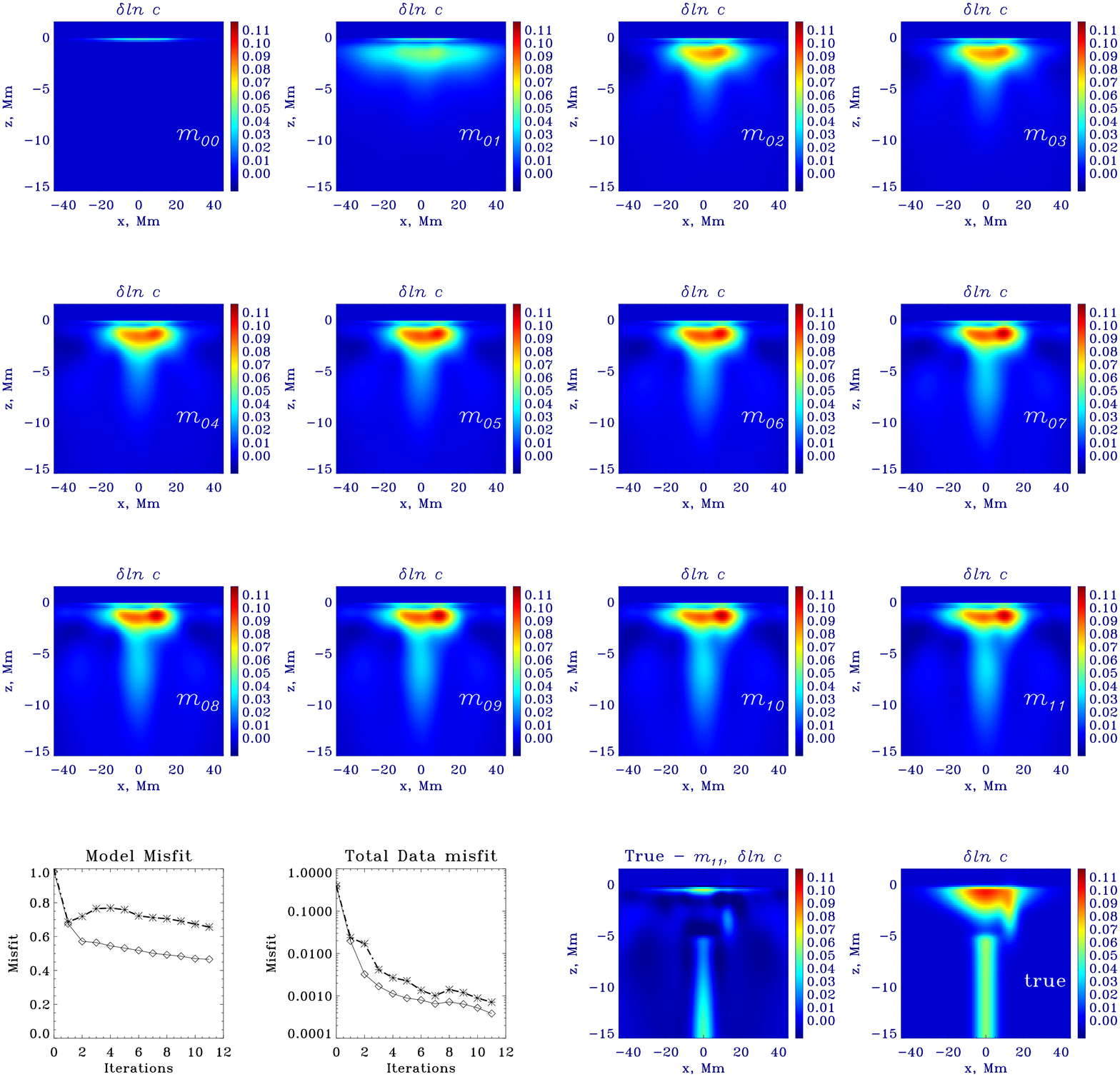}
\caption{Iterations in a conjugate-gradient based inversion. The starting model contains a `surface constraint', as seen in $m_{00}$. 
The rest of the algorithm is unchanged from the example shown in Figure~\ref{modelsconj}. The first iteration is performed using steepest descent and a combination
of conjugate gradient and steepest descent are used to compute subsequent models. It is seen that models approach the
true but the reduction in the misfit slows down with iteration. The model misfit is the normalized $L_2$ norm difference between 
the true and current model whereas the total data misfit is the same as equation~(\ref{misf}). For comparison, we over plot the misfit evolution
for the unconstrained inversion (dot-dashed line with asterisks). For categories of model and data misfit, it is seen that surface constraints
accelerate convergence.
\label{models.surface}}
\end{centering}
\end{figure}

\begin{figure}[!ht]
\begin{centering}
\epsscale{1}
\plotone{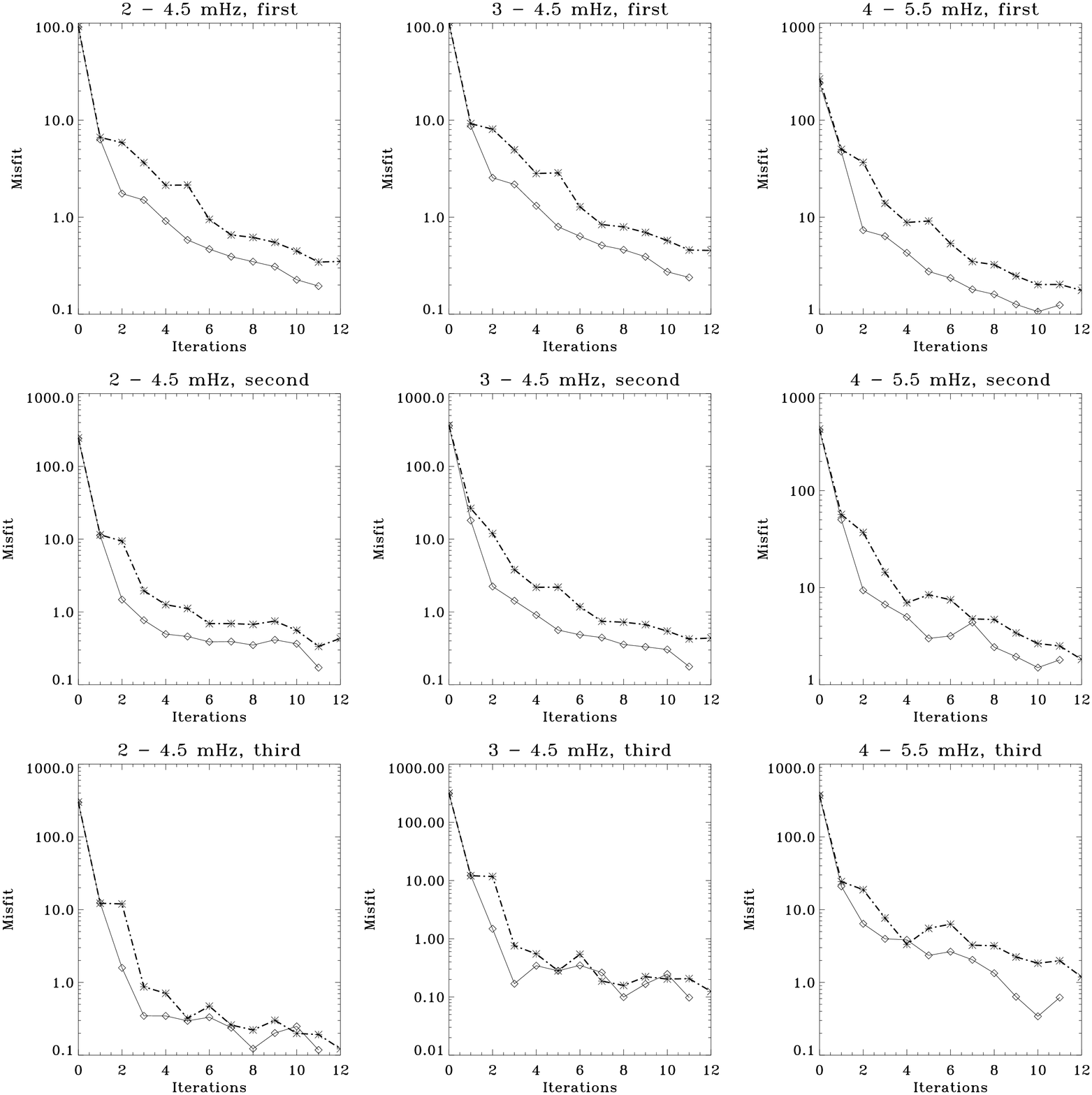}
\caption{Comparison of misfit bands between surface-constrained and unconstrained inversions. Systematically, unconstrained
inversions show slower convergence, as evidenced by the curves with higher misfit (dot-dashed lines with asterisk symbols).
Smooth lines with circle symbols show the misfit evolving with iteration for surface-constrained inversions.
\label{misfit.comp}}
\end{centering}
\end{figure}

In Figure~\ref{models.surface}, we see direct evidence of this, where the bottom-left panel shows a smooth decline in model misfit
with iteration, unlike in Figure~\ref{modelsconj}, which displays a non-monotonic trajectory. Overall, both data and model misfit
are lower in Figure~\ref{models.surface} in comparison to Figure~\ref{modelsconj}. We also over plot all the misfit categories in
Figure~\ref{misfit.comp} to highlight the (anticipated) superiority of surface-constrained inversions.

\begin{figure}[!ht]
\begin{centering}
\epsscale{1}
\plotone{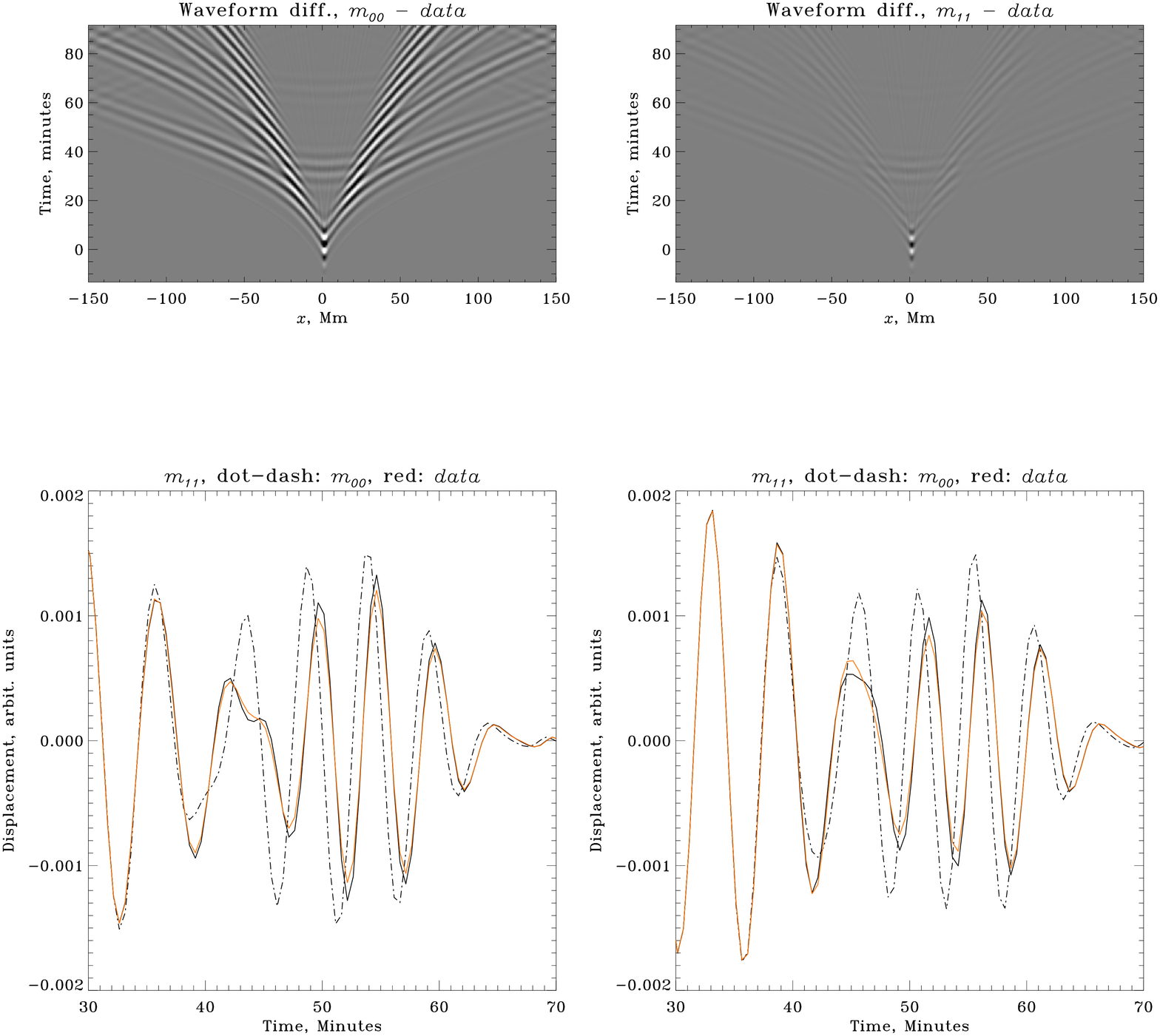}
\caption{Waveform matching as a function of iteration. difference between time-distance diagrams of
models $m_{00}$, $m_{11}$ and target data (upper panels). At iteration 11, the difference is substantially smaller (plotted on same scale).
Lower panels show waveforms at $x=-9$ Mm (left) and $x = 22$ Mm (right). By iteration 11, the waveforms match the data very well.
\label{waveform}}
\end{centering}
\end{figure}

Finally, we show the improvement between waveforms derived from ``data" and the model in Figure~\ref{waveform}. By iteration 11,
the waveforms start matching up well.

\section{Discussion}
Full waveform inversion provides a means of addressing longstanding problems in helioseismology. 
It directly addresses the major issue of non-linear
dependencies of travel times on properties of the solar medium in structures such as sunspots and supergranules. 
While iterative inversions are indeed possible using ray theory as the forward model, wave propagation is demonstrably 
not well captured in this high-frequency approximation \citep[][]{birch01}. Helioseismology is increasingly a high-precision 
science and to make accurate inferences, it is important to model wave effects as fully as possible. Finite frequency
forward calculations of the helioseismic wavefield are now routinely performed, and in this article we have discussed full waveform
inversion strategies within this context.

A basic lacuna of current approaches to 3-D helioseismic inversions is that there is rarely a consistency check of how much
the inverted model reduces the misfit between seismic prediction and observation. At each iteration in our inversion,
we perform a line search to determine how much to change model, and generally find that beyond 3-5\% the misfit actually
rises, suggesting that the linear connection between misfit and model change is restricted to this regime. Of course, the caveat
in drawing this conclusion is that our inversion method is either quasi-Newton- or conjugate gradient based, whereas
prior helioseismic inversions have relied on Gauss-Newton-based approaches. 
In general, Gauss-Newton allows for taking
larger steps in model space but it must be emphasized again that the actual extent to which misfit is reduced has generally
not been measured. 
The closest to a consistent inversion can be attributed to \citet{cameron08}, who attempted to
study a set of sunspot models using linear magneto-hydrodynamic numerical simulations to determine how well observations can be matched.
In a purely forward approach (``probabilistic"), the model space is exhaustively searched, determining the misfit for each
model. However, given the computational expense for full wave modeling codes, this may be an infeasible approach.

The methodology discussed here still requires development and a more careful exploration of techniques that can
enhance convergence. Purely computational test problems, such as the inversion for flows and magnetic fields, will
be the focus of future studies. However, full waveform inversion provides a firm theoretical
foothold for a field that has long sought a means to accurately interpret helioseismic measurements. The hope is that,
with the simultaneous development of inverse theory and  high-fidelity numerical methods to rapidly simulate wave propagation in a medium
that closely mimics the Sun, we may finally able to settle issues of great relevance to understanding solar dynamics.

\acknowledgements
  S.M.H. acknowledges funding from NASA grant NNX11AB63G. We also thank Hejun Zhu for his useful insights on FWI methodology.

\appendix
\section{Adjoint source}
We use equation~(4) from \citet{gizon_04} in order to define the weight function $W_{i}(t)$ for the travel-time measurement
\begin{equation}
W_{i}(t) = -{\dot{\mathcal C}^{\rm p}}_{i}(t)\,\frac{ f(t)}{\Delta t\sum_{t'}  f(t') \left[{\dot{\mathcal C}^{\rm p}}_{i}(t')\right]^2  }\,,\label{def.weight}
\end{equation}
where ${\mathcal C}^{\rm p}$ is the predicted waveform (cross correlation), $\Delta t$ is the temporal rate at which the waveform is sampled, $f(t)$ is a window, 
and the travel-time shift $\Delta\tau$ is given by
\begin{equation}
\Delta\tau_i = \int dt~ W_i(t)~({\mathcal C}_i^{\rm p} - {\mathcal C}_i^{\rm o}).
\end{equation}
The adjoint source is given by
\begin{equation}
f^\dagger({\bf x}, t) = \sum_i \Delta\tau_i\, W_i(-t)\, \delta({\bf x} - {\bf x}_i),
\end{equation}
where ${\bf x}_i$ is the a receiver (slave) and the summation is over all receivers.

\section{Steepest descent, Conjugate gradient and L-BFGS}
In all the methods described here, the model is updated thus, ${\bf m}^{k+1} = {\bf m}^{k} + \varepsilon{\bf p}^k$,
where $\varepsilon$ is obtained through a line search, i.e., $\varepsilon$ that minimizes $\chi({\bf m}^{k} +  \varepsilon{\bf p}^k)$.
Given the smoothed gradient at iteration $k$, ${\bf g}^k$.
The steepest descent update is simply ${\bf p}^k = -{\bf g}^k$. The conjugate gradient update
is given by 
\begin{equation}
{\bf p}^k = -{\bf g}^k + \beta^k\,{\bf p}^{k-1},\,\,\,\,\,\,\,\,\,\,\,\,\beta^k = \frac{{\bf g}^k\cdot({\bf g}^k - {\bf g}^{k-1})}{{\bf g}^k\cdot{\bf g}^k},
\end{equation}
and because there is a dependence on ${\bf p}^{k-1}$, the first iteration cannot also be performed by conjugate gradient.

The limited-memory BFGS update at iteration $N$ is obtained by manipulating the prior $m$ gradients and models.
The limited-memory aspect of this is accomplished by sweeping forward and reverse through prior gradients. 
\begin{eqnarray}
&&k = N\,\,\,\,\,\,\,\,\,\,\,{\bf h} = {\bf g}^k\nonumber\\
&&{\rm For\,\, k = N-1, N-2,...., N-m } \nonumber\\
&&\alpha^k = \frac{({\bf m}^k - {\bf m}^{k-1})\cdot{\bf h}}{({\bf m}^k - {\bf m}^{k-1})\cdot({\bf g}^k - {\bf g}^{k-1})} \nonumber\\
&&{\bf h} = {\bf h} - \alpha^k ({\bf g}^k - {\bf g}^{k-1}),\label{penul.eq}
\end{eqnarray} 

\begin{eqnarray}
&&{\rm For\,\, k = N-m, N-m+1,...., N-1 } \nonumber\\
&&\alpha^k = \alpha^k - \frac{({\bf g}^k - {\bf g}^{k-1})\cdot{\bf h}}{({\bf m}^k - {\bf m}^{k-1})\cdot({\bf g}^k - {\bf g}^{k-1})} \nonumber\\
&&{\bf h} = {\bf h} + \alpha^k ({\bf m}^k - {\bf m}^{k-1})\label{final.eq}
\end{eqnarray} 

The update is given by ${\bf p}^N = -{\bf h}$. The rule of thumb is to use between 3 and 7 prior gradients to construct 
the update, i.e., $3 \le m \le 7$ in equations~(\ref{penul.eq})~(\ref{final.eq}).


%

\bibliographystyle{apj}
\bibliography{../references}

\begin{thebibliography}{32}
\expandafter\ifx\csname natexlab\endcsname\relax\def\natexlab#1{#1}\fi

\bibitem[{{Birch} {et~al.}(2001){Birch}, {Kosovichev}, {Price}, \&
  {Schlottmann}}]{birch01}
{Birch}, A.~C., {Kosovichev}, A.~G., {Price}, G.~H., \& {Schlottmann}, R.~B.
  2001, \apjl, 561, L229

\bibitem[{{Bozda{\v g}} {et~al.}(2011){Bozda{\v g}}, {Trampert}, \&
  {Tromp}}]{bozdag11}
{Bozda{\v g}}, E., {Trampert}, J., \& {Tromp}, J. 2011, Geophysical Journal
  International, 185, 845

\bibitem[{{Cameron} {et~al.}(2008){Cameron}, {Gizon}, \& {Duvall}}]{cameron08}
{Cameron}, R., {Gizon}, L., \& {Duvall}, Jr., T.~L. 2008, \solphys, 251, 291

\bibitem[{{Christensen-Dalsgaard} {et~al.}(1996){Christensen-Dalsgaard},
  {Dappen}, {Ajukov}, {Anderson}, {Antia}, {Basu}, {Baturin}, {Berthomieu},
  {Chaboyer}, {Chitre}, {Cox}, {Demarque}, {Donatowicz}, {Dziembowski},
  {Gabriel}, {Gough}, {Guenther}, {Guzik}, {Harvey}, {Hill}, {Houdek},
  {Iglesias}, {Kosovichev}, {Leibacher}, {Morel}, {Proffitt}, {Provost},
  {Reiter}, {Rhodes}, {Rogers}, {Roxburgh}, {Thompson}, \& {Ulrich}}]{jcd}
{Christensen-Dalsgaard}, J., {Dappen}, W., {Ajukov}, S.~V., {Anderson}, E.~R.,
  {Antia}, H.~M., {Basu}, S., {Baturin}, V.~A., {Berthomieu}, G., {Chaboyer},
  B., {Chitre}, S.~M., {Cox}, A.~N., {Demarque}, P., {Donatowicz}, J.,
  {Dziembowski}, W.~A., {Gabriel}, M., {Gough}, D.~O., {Guenther}, D.~B.,
  {Guzik}, J.~A., {Harvey}, J.~W., {Hill}, F., {Houdek}, G., {Iglesias}, C.~A.,
  {Kosovichev}, A.~G., {Leibacher}, J.~W., {Morel}, P., {Proffitt}, C.~R.,
  {Provost}, J., {Reiter}, J., {Rhodes}, Jr., E.~J., {Rogers}, F.~J.,
  {Roxburgh}, I.~W., {Thompson}, M.~J., \& {Ulrich}, R.~K. 1996, Science, 272,
  1286

\bibitem[{{Dahlen} \& {Baig}(2002)}]{dahlen02}
{Dahlen}, F.~A., \& {Baig}, A.~M. 2002, Geophysical Journal International, 150,
  440

\bibitem[{{Dombroski} {et~al.}(2013){Dombroski}, {Birch}, {Braun}, \&
  {Hanasoge}}]{dombroski13}
{Dombroski}, D.~E., {Birch}, A.~C., {Braun}, D.~C., \& {Hanasoge}, S.~M. 2013,
  \solphys, 282, 361

\bibitem[{{Duvall} {et~al.}(1993){Duvall}, {Jefferies}, {Harvey}, \&
  {Pomerantz}}]{duvall}
{Duvall}, Jr., T.~L., {Jefferies}, S.~M., {Harvey}, J.~W., \& {Pomerantz},
  M.~A. 1993, \nat, 362, 430

\bibitem[{{Fichtner} {et~al.}(2009){Fichtner}, {Kennett}, {Igel}, \&
  {Bunge}}]{fichtner09}
{Fichtner}, A., {Kennett}, B.~L.~N., {Igel}, H., \& {Bunge}, H.-P. 2009,
  Geophysical Journal International, 179, 1703

\bibitem[{{Giles} \& {Pierce}(2000)}]{giles.adjoint.00}
{Giles}, M.~B., \& {Pierce}, N.~A. 2000, Flow, Turbulence and Combustion, 65,
  393

\bibitem[{{Gizon} \& {Birch}(2002)}]{gizon02}
{Gizon}, L., \& {Birch}, A.~C. 2002, \apj, 571, 966

\bibitem[{{Gizon} \& {Birch}(2004)}]{gizon_04}
---. 2004, \apj, 614, 472

\bibitem[{{Gizon} \& {Birch}(2005)}]{gizon05}
---. 2005, Living Reviews in Solar Physics, 2, 6

\bibitem[{{Gizon} {et~al.}(2010){Gizon}, {Birch}, \& {Spruit}}]{gizon2010}
{Gizon}, L., {Birch}, A.~C., \& {Spruit}, H.~C. 2010, \araa, 48, 289

\bibitem[{{Hanasoge} {et~al.}(2012){Hanasoge}, {Birch}, {Gizon}, \&
  {Tromp}}]{hanasoge12_mag}
{Hanasoge}, S., {Birch}, A., {Gizon}, L., \& {Tromp}, J. 2012, Physical Review
  Letters, 109, 101101

\bibitem[{{Hanasoge}(2007)}]{hanasoge_thesis}
{Hanasoge}, S.~M. 2007, PhD thesis, Stanford University, California, USA

\bibitem[{{Hanasoge} {et~al.}(2011){Hanasoge}, {Birch}, {Gizon}, \&
  {Tromp}}]{hanasoge11}
{Hanasoge}, S.~M., {Birch}, A., {Gizon}, L., \& {Tromp}, J. 2011, \apj, 738,
  100

\bibitem[{{Hanasoge} {et~al.}(2008){Hanasoge}, {Couvidat}, {Rajaguru}, \&
  {Birch}}]{Hanasoge_couvidat_2008}
{Hanasoge}, S.~M., {Couvidat}, S., {Rajaguru}, S.~P., \& {Birch}, A.~C. 2008,
  \mnras, 391, 1931

\bibitem[{{Hanasoge} \& {Duvall}(2007)}]{dealias}
{Hanasoge}, S.~M., \& {Duvall}, Jr., T.~L. 2007, Astronomische Nachrichten,
  328, 319

\bibitem[{{Hanasoge} {et~al.}(2010){Hanasoge}, {Komatitsch}, \&
  {Gizon}}]{hanasoge_2010}
{Hanasoge}, S.~M., {Komatitsch}, D., \& {Gizon}, L. 2010, \aap, 522, A87

\bibitem[{Hu {et~al.}(1996)Hu, Hussaini, \& Manthey}]{hu}
Hu, F.~Q., Hussaini, M.~Y., \& Manthey, J.~L. 1996, Journal of Computational
  Physics, 124, 177

\bibitem[{{Jackiewicz} {et~al.}(2012){Jackiewicz}, {Birch}, {Gizon},
  {Hanasoge}, {Hohage}, {Ruffio}, \& {{\v S}vanda}}]{jason12}
{Jackiewicz}, J., {Birch}, A.~C., {Gizon}, L., {Hanasoge}, S.~M., {Hohage}, T.,
  {Ruffio}, J.-B., \& {{\v S}vanda}, M. 2012, \solphys, 276, 19

\bibitem[{Lele(1992)}]{lele92}
Lele, S.~K. 1992, Journal of Computational Physics, 103, 16

\bibitem[{{Luo} {et~al.}(2013){Luo}, {Modrak}, \& {Tromp}}]{tromp13}
{Luo}, Y., {Modrak}, R., \& {Tromp}, J. 2013, in Handbook of geomathematics,
  2nd edn., ed. W.~Freeden, M.~Z. Nashed, \& T.~Sonar (Springer Verlag)

\bibitem[{Rickers {et~al.}(2013)Rickers, Fichtner, \& Trampert}]{rickers2013}
Rickers, F., Fichtner, A., \& Trampert, J. 2013, Earth and Planetary Science
  Letters, 367, 39

\bibitem[{{Snieder}(2004)}]{snieder04}
{Snieder}, R. 2004, \pre, 69, 046610

\bibitem[{{Stein} \& {Nordlund}(2000)}]{stein00}
{Stein}, R.~F., \& {Nordlund}, {\AA}. 2000, \solphys, 192, 91

\bibitem[{Tape {et~al.}(2009)Tape, Liu, Maggi, \& Tromp}]{CarlTape09}
Tape, C., Liu, Q., Maggi, A., \& Tromp, J. 2009, Science, 325, 988

\bibitem[{{Tarantola}(1984)}]{tarantola84}
{Tarantola}, A. 1984, {Geophysical Prospecting}, 32, 998

\bibitem[{{Tromp} {et~al.}(2010){Tromp}, {Luo}, {Hanasoge}, \&
  {Peter}}]{tromp10}
{Tromp}, J., {Luo}, Y., {Hanasoge}, S., \& {Peter}, D. 2010, Geophysical
  Journal International, 183, 791

\bibitem[{{Tromp} {et~al.}(2005){Tromp}, {Tape}, \& {Liu}}]{tromp05}
{Tromp}, J., {Tape}, C., \& {Liu}, Q. 2005, Geophysical Journal International,
  160, 195

\bibitem[{{{\v S}vanda} {et~al.}(2011){{\v S}vanda}, {Gizon}, {Hanasoge}, \&
  {Ustyugov}}]{svanda11}
{{\v S}vanda}, M., {Gizon}, L., {Hanasoge}, S.~M., \& {Ustyugov}, S.~D. 2011,
  \aap, 530, A148

\bibitem[{Zhu {et~al.}(2013)Zhu, Bozda{\u g}, Duffy, \&
  Tromp}]{zhu_attenuation}
Zhu, H., Bozda{\u g}, E., Duffy, T.~S., \& Tromp, J. 2013, Earth and Planetary
  Science Letters, 381, 1

\end{thebibliography}

\end{document}